# Step Detection Algorithm For Accurate Distance Estimation Using Dynamic Step Length

Ahmad Abadleh
IT Department
Mutah University
Mutah, Karak, Jordan
ahmad_a@mutah.edu.jo

Eshraq Al-Hawari
IT Department
Mutah University
Mutah, Karak, Jordan
eshraqh@mutah.edu.jo

Esra'a Alkafaween
IT Department
Mutah University
Mutah, Karak, Jordan
Esra_ok@mutah.edu.jo

Hamad Al-Sawalqah
Computer Information Systems Department
The University of Jordan
Amman, Jordan

*Abstract*— In this paper, a new Smartphone sensor based algorithm is proposed to detect accurate distance estimation. The algorithm consists of two phases; the first phase is for detecting the peaks from the Smartphone accelerometer sensor. The other one is for detecting the step length which varies from step to step. The proposed algorithm is tested and implemented in real environment and it showed promising results. Unlike the conventional approaches, the error of the proposed algorithm is fixed and is not affected by the long distance.

*Keywords*— distance estimation; peaks; step length; accelerometer.

## I. Introduction

Recently, Smartphones have become the most common computing devices that are in widespread use in our daily lives around the world. During the last decade, many improvements have been added to the Smartphones, including the number of communication modules (e.g. GPS) and sensors modules (e.g. accelerometer, gyroscope, digital compass, and magnetometer sensor) that are embedded in the device [1]. By using these modules, many Smartphones services and applications have been developed (e.g. indoor/outdoor localization) [1, 2, 11]. Indoor location services have been gaining an increasing focus in recent years.

Pedestrian dead reckoning (PDR) is one of well known techniques for indoor positioning using Smartphone sensors. PDR is a technique to estimate the latest user position using the embedded inertial sensors by combining a known position with the successive position displacements (distance) [3]. The user walking distance can be estimated by double integration of the accelerations with respect to time. However, due to the noise in the accelerometer output, any error in the signal will accumulate rapidly with time. To avoid this error, the distance is estimated by multiplying the number of steps by a constant step length [2, 4].

User's step and step length are important issues in estimating the walking distance. Several methods have been proposed for the detection of the steps, such as, detection of the peaks [5], flat zone detection [6] and zero-crossings method [4, 7]. Many previous works, e.g., [8, 9] proposed different methods to estimate the step length; some of them used static length of each step, while others used dynamic step length [2]. Most of the current approaches suffer from the unpredictable movements of the users which affect the accuracy of the results. For instance, using fixed step length and the error in counting the steps are the major problems of the step detection algorithms.

In this paper, a new step detection algorithm is presented to enhance the performance of the travelled distance estimation using dynamic step length. The proposed algorithm utilizes the acceleration values of X, Y and Z axes from accelerometer sensor, which is built in Smartphone to count steps in real-time. In our work, we assume that the mobile phone is in a static position placed in user's hand throughout the movements. The proposed algorithm determines the peaks precisely and a weight is assigned to each step based on its length. To evaluate our algorithm, real experiments were conducted and a comparison with the conventional methods were presented.

The proposed algorithm uses a new technique to count the peaks in which the miscounting peaks are reduced. Moreover, it detects the length of the step and categorizes them into different categories which lead to better distance estimation. The main contributions of this paper are:

- Detecting the length of the steps.
- Categorizing the steps into short, medium, and long step.
- Mitigating the problem of counting the peaks due to irrelevant movements.

The rest of this paper is organized as follows: Section II provides some related works about distance estimation using mobile phones and methods for step detection. Section III introduces the overall architecture of our system for travelled distance estimation, and then presents our algorithm for peaks and step length detecting. Details about the experiments and the results are presented in Section IV. Finally, the paper ends with conclusion and future works in Section V.

## II. Related work

Various methodologies and techniques have been proposed and evaluated for user distance estimation [1 – 5]. Researchers have presented different approaches based on fixed step length, and others have used dynamic step length for distance estimation.

Liu et al. presented a novel algorithm for adaptive step length estimation using a handheld device based on empirical formula and Back-propagation neural network. They relied on tri-axial accelerometer to detect the steps using hamming window function with digital low-pass filter to tri-axial accelerometer [8].

Shih et al. developed a waist-mounted based PDR system to estimate the horizontal walking distance using Pythagoras' theorem from the height change of the waist. They used vertical acceleration to detect the step and the Pythagoras' theorem to estimate each step length [9].

Zeng et al. relied on their work on the acceleration and gravity sensor to count the steps, where two different thresholds used for the detection of peaks and valleys, and three adjacent points were stored in the memory. The comparison between the two points based on the altitude and time. if there are two adjacent points and one of them is the peak and the other is valley, then there is a possibility of a new step. But if the points are all either peaks or valleys, the small point is deleted. If the period of time between two adjacent points is very short, then the point is deleted as well [3].

Another mobile phone based approach have been presented by Alzantot.et al, where they have suggested to use a finite state machine approach combining with threshold values settings for the acceleration measurements to estimate user steps. They also employed a Support Vector Machine classifier for estimating the varying stride length of the user [1].

A key observation drawn from the literatures related to travelled distance estimation is that some of the studies are based on fixed step length and others have used dynamic step length in their approaches [1, 3, 8, 9, 10].

Our works differ from the current works as follows:

- Fixed step length based approaches suffer from the accumulative error in distance estimation. However, our approach uses the dynamic step length to mitigate that error.
- Dynamic step length based approaches need many empirical thresholds values, such as: peak threshold, constant values to detect the length. Our approach categorizes the length into different categories and assigns a weight for each step.
- Current peak detection algorithms have the problem of counting the peaks due to the irrelevant movements. Our approach reduces this problem by using the proposed MAX approach.

### III. SYSTEM ARCHTECTURE

The proposed approach has the following sections: data collection phase, filtering phase, peak detection phase, and distance estimation phase.

#### A. Data collection phase

During this phase, the three axis accelerometer data is derived from the Smartphone assuming that the user holds the Smartphone by hand. The derived data includes X, Y, and Z values. Then the magnitude value is computed as the square root of the three accelerometer values as shown in Eq. 1. The magnitude value represents the vibration on the three axis coordinates.

$$mag = \sqrt{x^2 + y^2 + z^2} \quad (1)$$

In order to exclude the gravity force from the magnitude value, the average of the magnitude values, which is computed from (1), is computed. Then each magnitude (*netmag*) value is computed as the difference between the current magnitude (*mag*) and average magnitude value (*avgmag*) as shown in (2).

$$netmag = mag - avgmag \quad (2)$$

Fig. 1 shows an example of the collected data on the three axes, while Fig. 2 shows the magnitude value which is computed by (1).

#### B. Filtering phase

The Smartphone sensors are inaccurate and suffer from various problems. Therefore, the derived data must be filtered in order to exclude the noise and the outliers' values. Smoothing the data helps in improving the accuracy of the system as the accelerometer sensor is very sensitive to the movements. The process of choosing peaks will be easier and accurate if the data is smoothed and has a standard shape.

During this phase, kalmn filter is applied to help in excluding the gravity force. While high pass filter is used to remove the outliers' values. The data is smoothed by performing the filters which enhanced the result.

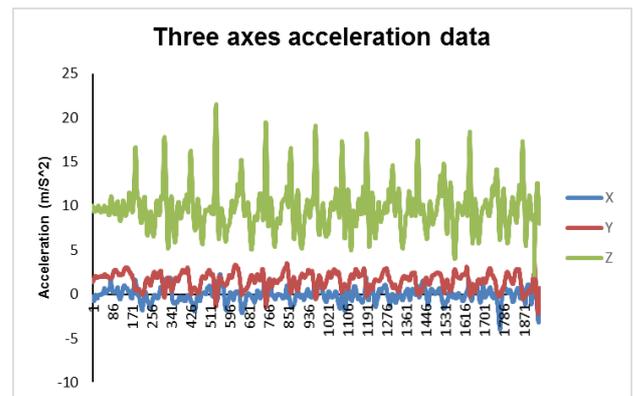

Fig. 1   Acceleration on three axes

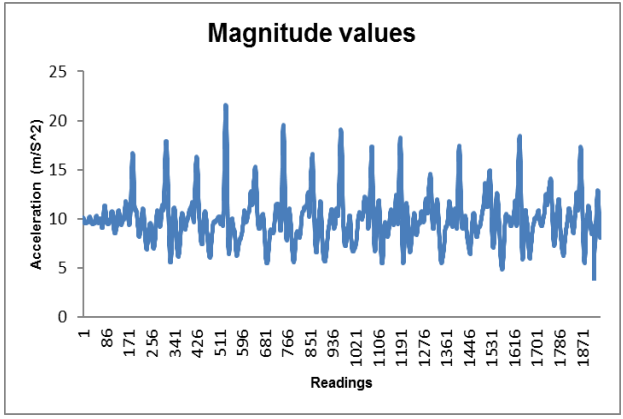

Fig. 2   Magnitude values

## C. Peak detection phase

During the user movements, the changes on the acceleration can be useful. For instance, z-axis gets affected by the vertical movement such as raising the leg. Therefore, this particular change forms some peaks on the acceleration. Exploiting these peaks can help in distance estimation. Most of the current step detection algorithms rely on counting the peaks as each peak represents a user stride. The distance is computed by multiplying the number of steps by a step length. However, step length varies even for the same user; therefore, fixed step length can lead to an error in distance estimation. This error can be severe if the walking distance is long. Another important problem is the fake peaks. Fake peak represents a peak produced by an irrelevant movement. Fig. 3 shows the real and the fake peaks. Counting the fake peak as a real peak increases the error in the distance estimation. Since the peak is detected if the acceleration surpasses a specific threshold, fake peak might be counted as a real peak as shown in Fig. 3.

In this paper, a new algorithm is proposed to detect the real peaks and the step length. The algorithm has two stages, in the first stage; the peak is detected while the second stage is responsible about detecting the length of the steps.

- **Peak detection**: the algorithm works as follows: the time is divided into dynamic time windows in which each time window contains the magnitude values which are computed by (2). The algorithm keeps track of three consecutive values in a vector called Start Vector, if the three values getting increased, then a peak might be a head. Then the first value represents the start point of the peak. If the value surpasses a step threshold which is defined in advance, then all the values above the threshold will be stored in a vector called Peak Vector. The algorithm utilizes the peak vector to detect the real peak values as well as the time when they are taken place. Since the peak vector contains the values that fall above the step threshold, then the maximum value should represent the real peak.

- **Fake peak Step length detection**: after the peak is detected, the signals should be decreased as the peak is the maximum value. The algorithm keeps track of two consecutive values in a vector called *End Vector*. The end point of the peak is detected when the two values are getting increased, which means that the step has finished. The step length is computed as the difference between the end and the start points.

Peak vector mitigates the error that can be happened due to the fake peaks as only one peak (max value) will be chosen. Moreover, dynamic step length makes the algorithm more robust to the change of the user walking style. Fig. 4 illustrates the main idea of the algorithm.

## D. Distance estimation phase

In previous section, the two important variables, the peak and the step length are detected. In this section, the step length is categorized into three categories, short which means less than the average, medium which means equal to the average, and long step which means greater than the average. Each category is assigned a weight to represent its length as shown in (3).

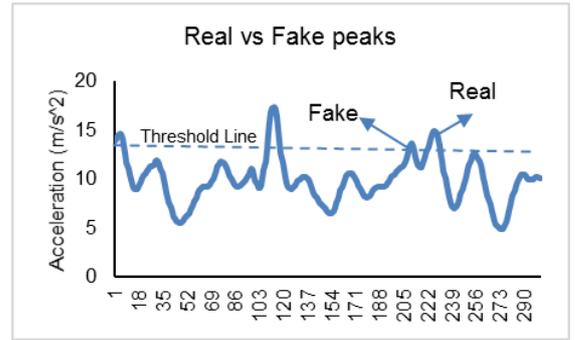

Fig. 3.   Real vs Fake peaks

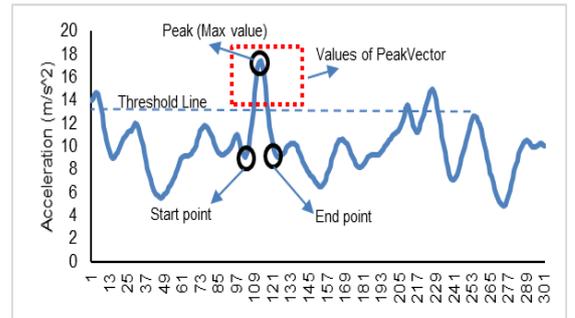

Fig. 4.   Illustration of the proposed algorithm

$$\alpha = \begin{cases} 0.5 & if\ StepSize < Mag_{avg} \\ 1.0 & if\ StepSize = Mag_{avg} \\ 1.5 & if\ StepSize > Mag_{avg} \end{cases} \quad (3)$$

The distance then is estimated using (4):

$$Distance = \sum_{i=1}^{n} Peak(i) * \alpha(i) \quad (4)$$

Equation. 4 represents the distance estimation where the length of the step is taken into account. The result of the distance estimation shows promising results and there is enhancement in distance estimation as the error is reduced. Furthermore, unlike the conventional step detection algorithms which have accumulative error, the error of this algorithm is fixed even if the distance is long (more details in the evaluation section).

## IV. EVALUATION AND RESULTS

Our proposed approach is tested in a real environment at the department of computer science in Mutah University. In this section, the results of the proposed approach are presented.

### A. Environmental experiments

The user walked on the corridors at the department and passed different distances. During the walking, the user changed the speed from time to time and walked on a straight line and sometimes on a winding line.

### B. Experiment results

The result of the experiments is shown in Table 1 and Fig. 5. As can be seen in Table 1, in different distances, the proposed approach performs well and has a good accuracy.

TABLE1. PROPOSED APPROACH RESULTS

| Gender | Real Distance /m | Est. number of steps | Est. distance /m |
|---|---|---|---|
| Male | 10 | 9 | 9.6 |
| Male | 20 | 19 | 19.5 |
| Male | 35 | 36 | 33.3 |
| Male | 100 | 105 | 103 |
| Female | 10 | 10 | 9.2 |
| Female | 20 | 24 | 23 |
| Female | 35 | 34 | 33 |
| Female | 100 | 101 | 98 |

Fig. 5 shows that the estimated distance is almost identical with real one. The reason behind that is because the proposed approach outperforms the problem of fake peaks and the difference on the length of each step. It can be seen that the proposed approach is robust even if the distance is long. The algorithm guarantees the scalability as the error is almost fixed for all the distances. The accuracy of the proposed approach is better than the one with conventional method and comparable with the approaches with dynamic step length.

Fig. 6 represents the comparison between the conventional peak detection algorithm in which constant step length is used and the proposed approach. It reinforces our claim that the conventional approach suffers from the accumulative error as the distance is getting increased, while the proposed approach has almost a fixed error. From Fig. 6, the average error of the proposed approach was 1.4 m while the conventional one has an average error of 4.8 m.

Applying the proposed approach mitigated the problem of irrelevant movement, such as hand shaking or any sudden change during the user movement. The idea of getting the maximum value among the values that fall above the threshold reduces the influence of the hand shaking problem.

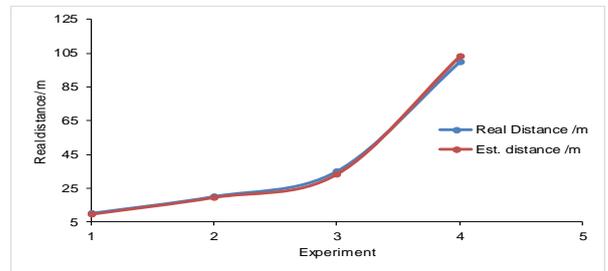

Fig. 5. Estimated distance using the proposed approach compared to the real one.

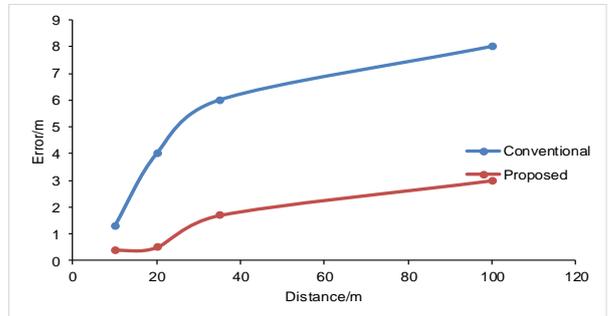

Fig. 6. The proposed approach results compared to conventional peak detection algorithm.

## V. CONCLUSION

Distance estimation is one of the critical issues in indoor localization in which it affects the accuracy. In this paper, a new approach has been proposed in order to detect accurate and precise distance. The proposed approach enhanced the distance estimation by taking into account the step length and the error in counting the peaks, which are not solved perfectly in most of the existing approaches.

In this work, a new technique for detecting the step length was proposed and an enhanced peak detection algorithm was proposed as well. The results of the real experiments showed enhancement in distance estimation and the error became fixed on different experiments. According to the results, we claim that the proposed approach is robust and effective. Most of the distance estimation approaches face the handshaking problem which causes degradation in the accuracy. In our approach, the problem is mitigated but not perfectly solved. In the future work, we will consider the handshaking problem as well as holding the Smartphone in any place.